# Spin Drift in Highly Doped n-type Si

Makoto Kameno[1], Yuichiro Ando[1], Teruya Shinjo[1], Hayato Koike[2],
Tomoyuki Sasaki[2], Tohru Oikawa[2], Toshio Suzuki[3] and Masashi Shiraishi[1, 4]

1. Graduate School of Engineering Science, Osaka Univ. Osaka, Japan
2. Advanced Technology Development Center, TDK Cooperation, Chiba, Japan
3. AIT, Akita Research Institute of Advanced Technology, Akita, Japan
4. Graduate School of Engineering, Kyoto Univ. Kyoto, Japan.

Corresponding author: M. Shiraishi (mshiraishi@kuee.kyoto-u.ac.jp)

**Abstract**

A quantitative estimation of spin drift velocity in highly doped n-type silicon (Si) at 8 K is presented in this letter. A local two-terminal Hanle measurement enables the detection of a modulation of spin signals from the Si as a function of an external electric field, and this modulation is analyzed by using a spin drift-diffusion equation and an analytical solution of the Hanle-type spin precession. The analyses reveal that the spin drift velocity is linearly proportional to the electric field. The contribution of the spin drift effect to the spin signals is crosschecked by introducing a modified nonlocal four-terminal method.

Spin transport in a semiconductor has two attributes: spin diffusion and spin drift. Spin diffusion is caused by a gradient of spin densities for up-spins and down-spins, and spin drift is induced by an electric field in the spin channel. Generation of a pure spin current and its transport under spin diffusion were achieved in nonmagnetic materials by using an electrical nonlocal four-terminal (NL-4T) scheme [1]. This method has the potential to demonstrate spin diffusion in semiconductors as well [2-6], and has been widely used in semiconductor spintronics. In Si spintronics, spin injection and transport of a pure spin current has been realized in highly doped n-type Si by using the NL-4T method, and estimation of spin coherence under spin diffusion has been achieved by Hanle effect measurements [2,7,8]. While spin diffusion took place in the NL-4T method, spin drift made a significant contribution to spin transport in a semiconductor [9,10]. It was theoretically shown that the spin drift induced by an electric field in a semiconductor spin channel enables modulation of the spin transport length scale [9,10], and in fact, spin drift was experimentally manifested in GaAs [11,12] and graphene [13]. In Si spintronics, the spin transport length scale in Si was experimentally modulated by using a nonlocal three-terminal (NL-3T) method when an electric field was applied in a part of the spin detection circuit [14,15]. However, the spin drift velocity in Si has yet to be qunatified.

The spin polarized current, $j_s$, is defined as the difference between the up-spin flow, $j_\uparrow$, and down-spin flow, $j_\downarrow$:

$$j_s = j_\uparrow - j_\downarrow = e(n_\uparrow - n_\downarrow)\mu E + eD\nabla(n_\uparrow - n_\downarrow), \qquad (1)$$

where $\mu$ is the spin mobility, $E$ is the electric field, $e$ is the elementary charge, $D$ is the spin diffusion constant, and $\nabla n_{\uparrow(\downarrow)}$ is the gradient of the up (down) carrier density. The $\mu E$

(=$v$) in this equation represents the spin drift velocity (in the case of charge flow, $\mu E$ represents charge drift velocity). As seen in Eq. (1), the spin polarized current is described as the sum of spin drift and spin diffusion terms. In order to study the contribution of spin drift in spin transport, an electric field must be applied to the semiconductor spin channel. Thus, the NL-4T method cannot be used because no electric field is applied in the signal detection circuit. In the present study, we conduct a quantitative investigation of spin drift velocity in a highly doped Si by analyzing spin signals. We introduce modified nonlocal and local methods, whereby an electric field can be applied in the Si spin channel. For these experiments, we use a Si spin valve, through which room-temperature spin transport was confirmed by NL-4T in order to avoid the possible misinterpretation of results that can occur in the nonlocal 3-terminal method [16]. The relationship between spin signals and spin drift velocity is also discussed on the basis of the spin drift-diffusion model.

Figure 1(a) illustrates the structure of a sample consisting of phosphorus-doped n-type Si (~$5\times10^{19}$ cm$^{-3}$: degenerated Si) on a silicon-on-insulator (SOI) substrate. The width and thickness of the channel are 21 μm and 80 nm, respectively. The device was equipped with two ferromagnetic (FM) electrodes consisting of Fe-on-MgO tunneling barrier layers (0.8-nm-thick) in order to avoid a conductivity mismatch. The conductivity of Si was measured to be $1.38\times10^{5}$ Ω$^{-1}$m$^{-1}$ at 8 K. The nonmagnetic (NM) electrodes were made of Al, and thus the junction between the NM electrodes and Si is ohmic. Figure 1(b) shows the spin signals obtained by using the NL-4T method under a dc bias current at room temperature. This result clearly demonstrates the successful spin transport in our device, allowing us to investigate spin transport properties precisely. All experiments in the present study were performed at 8 K in order to reduce the noise, and the spin diffusion constant, spin lifetime and spin diffusion length at 8 K were estimated to be 7.3 cm$^2$/s, 6.5 ns and 2.2 μm, respectively. These estimates were made by averaging the spin signals obtained by using NL-4T Hanle measurements (see Fig. 1(c)).

The experimental setup we used to investigate the spin drift velocity in the highly doped n-type Si is shown in Fig. 2(a). We employed a two-terminal (2T) method, using one FM and one NM electrode in order to measure the spin accumulation signals beneath the FM. Since an electric field is applied in the entire Si channel, this method enables to investigate the effect of spin drift in the highly doped Si in this setup. A Hanle-type spin precession of the spins was induced beneath the FM by applying a perpendicular magnetic field to the plane of the Si channel in order to estimate the spin drift velocity. Although background voltages were observed, these voltages did not depend on the perpendicular magnetic field and were subtracted in the analyses. The measurements of the 2T Hanle effect were repeated several times in order to obtain a sufficient signal-to-noise ratio. In the presence of both spin drift and spin diffusion in the Si channel, spin transport is described by the following spin drift-diffusion equation:

$$\frac{\partial S}{\partial t} = D\frac{\partial^2 S}{\partial x^2} + v\frac{\partial S}{\partial x} - \frac{S}{\tau}, \qquad (2)$$

where $S(x,t)$ is equal to $n_\uparrow - n_\downarrow$, $v$ is the spin drift velocity, $x$ is position, $t$ is time, and $\tau$ is the spin lifetime. An analytical fitting function for the Hanle effect is derived from Eq. (2) by using Green's function and taking into account spin precession. The analytical fitting function is given by:

$$\frac{V(B)}{I} = \pm \frac{P^2 \sqrt{DT}}{2\sigma A} \exp\left(-\frac{L}{\lambda_N} + \frac{L}{2\lambda_N^2} v\tau\right)(1+\omega^2 T^2)^{-1/4}$$

$$\times \exp\left[\frac{-L}{\lambda_N}\left(\sqrt{\frac{\sqrt{1+\omega^2 T^2}+1}{2}} - 1\right)\right]$$

$$\times \cos\left[\frac{\sqrt{(\text{Tan}^{-1}(\omega T))^2}}{2} + \frac{L}{\lambda_N}\sqrt{\frac{\sqrt{1+\omega^2 T^2}-1}{2}}\right], \quad (3)$$

where $P$ is spin polarization, $A$ is the cross-sectional area of the channel, $L$ is the gap length between two FM electrodes, $\omega = g\mu_B B/\hbar$ is the Larmor frequency, $g$ is the $g$-factor of an electron ($g = 2$ in this study), $\mu_B$ is the Bohr magneton, $\hbar$ is the Dirac constant, and the spin diffusion length can be expressed as $\lambda_N = \sqrt{D\tau}$. The derivation of Eq. (3) is described in detail in Refs. 17 and 18. Note that $T^{-1}$ is given by $T^{-1} = v^2/4D + 1/\tau$ under spin drift. The distance, $L$, was set to zero for the 2T case. The bias dependence of spin signals under the Hanle effect was measured by applying a dc bias current of -1.0, -1.5, -2.0, -3.0, and -4.0 mA. Here, a negative bias current is defined as the case when spins are extracted from the Si channel.

Obvious Hanle signals were observed in the 2T method at 8 K under each condition, as shown in Fig. 2(b). The spin drift velocity under each bias condition was calculated by using Eq. (3); the results are shown in Fig. 2(c) (the horizontal axis of Fig. 2(c) is the electric field calculated from the conductivity of the Si channel). Notably, the estimated spin drift velocity is almost proportional to the electric field. The spin drift velocity, $v$, is defined as the product of the mobility, $\mu$, and the electric field, $E$. The results shown in Fig. 2(c) are nicely explained by this relationship. The mobilities estimated from the conductivity of Si and from the slope of the fitting line in Fig. 2(c) are ca. 170 and 300 cm$^2$/Vs, respectively. Here, the relationship between the mobility and diffusion constant is linear for both charge and spin, and the discrepancy between the charge diffusion constant, $D_c$, and the spin diffusion constant, $D_s$, can induce a discrepancy between the spin mobility and charge mobility. While the cause of the discrepancy between the charge and spin mobilities is still unclear and further study is necessary, we conclude that this may be attributed to the difference between $D_c$ and $D_s$ as observed in other materials [19, 20]. Maassen *et al.* estimated the $D_c$ of single-layer graphene (SLG) by a Hall conductivity measurement and its $D_s$ by the Hanle measurement, and found that $D_c$ was 50-80 times greater than $D_s$ in SLG [19]. A similar suppression of the spin diffusion constant ("spin Coulomb drag") was reported in a two-dimensional electron gas system in GaAs/AlGaAs [20]. In that study, the spin and charge mobilities of Si were estimated by two different approaches, namely, measurements of the Hanle effect and charge conductivity, as in the case of SLG. Thus, while our results may point to a difference between $D_c$ and $D_s$, the discrepancy in our study is rather small and the tendency is the opposite in Si.

Next, we discuss the relationship between the spin signal and spin drift velocity by focusing on an analytical fitting function by taking into account the effect of spin drift (Eq. (3)), where the spin signal varies as a function of the spin drift velocity. Substituting $B=0$ into Eq. (3), the spin resistance can be expressed as,

$$R_s = \frac{V_{B=0}}{I} = \pm \frac{P^2 \sqrt{DT}}{2\sigma A} \exp\left[-\frac{L}{\lambda_N}\left(1 - \frac{v\tau}{2\lambda_N}\right)\right]. \quad (4)$$

Hence, the ratio of the spin resistance with and without spin drift can be written as:

$$\frac{R_s \text{ with drift}}{R_s \text{ w/o drift}} = \sqrt{\frac{T}{\tau}} \exp\left[\frac{v\tau}{2\lambda_N^2}L\right] = \frac{\exp\left[\frac{L}{2D}v\right]}{\sqrt{1 + \frac{\lambda_N}{2D}v}}. \quad (5)$$

Figure 3(a) shows a simulated curve of the relationship between the spin drift velocity and the ratio of the spin resistance with and without spin drift. This ratio is modulated by the spin drift velocity when $L$ is not equal to zero, implying that spin drift has a measurable effect on the detection of spin signals. In order to confirm the relationship in Eq. (5), we conducted the following experiment. Figures 3(b) and 3(c) show the experimental setups for the conventional NL-4T method and a modified NL-4T ("crossed NL-4T") method, respectively. An electric field is applied in the spin transport channel between the FM electrodes in the crossed NL-4T method. The spin drift velocity in the channel can be controlled by adjusting the electric field, i.e., the bias electric current. Figure 3(d) compares spin signals as a function of bias current in the conventional and crossed NL-4T methods. Note that the bias dependence of the spin signals in the crossed method was clearly different from those in the conventional method. Under an applied positive (negative) bias current, the spin signals in the crossed method are enhanced (suppressed), although intrinsic spin transport parameters such as the spin diffusion length is constant for all bias conditions [4]. The experimental results of the Hanle signals obtained in the crossed NL-4T method are shown in Fig. 3(e), where the red and blue open circles denote the Hanle signals at 8 K, obtained under a dc bias current of -4 mA and +4 mA, respectively. The black open circles show the Hanle signals obtained by the conventional NL-4T method under a dc bias current of +4 mA As seen in Fig. 3(e), the spin signals were clearly affected by the magnitude and polarity of the bias current, and the signal intensity under a positive dc bias current was about seven times larger than that under a negative dc bias current, because the spin transport length scale was enhanced owing to the spin drift. The supporting evidence is shown in Fig. 3(f), where the simulated curves calculated from Eq. (5) are plotted. These theoretical curves can accurately reproduce the experimentally obtained curves. In the theoretical curves, the spin polarization, the spin diffusion constant, the spin lifetime, the spin diffusion length were set to 0.04, 7.3 cm$^2$/s, 6.5 ns and 2.2 μm, respectively. In addition, the magnitude of the spin drift velocity was estimated to be 511 m/s as shown in Fig. 2(c). The agreement corroborates our quantitative estimation of the spin drift velocity.

In summary, we investigated the effect of spin drift by applying an electric field in a highly doped Si channel, and found that the spin drift velocity exhibited a linear dependence on the electric field. Our spin drift velocity estimate was comparable to the product of the Si mobility and the electric field in the Si channel. A "crossed" NL-4T method was introduced, where the direction and the magnitude of the spin drift velocity was controlled as a function of the bias current. A modulation of the spin signals due to the spin drift was demonstrated at 8 K, where the magnitude of the spin signal under +4 mA was about seven times larger than that under -4 mA. This strong enhancement corroborates that the spin drift strongly affects the magnitude of the spin signals in Si.

**Figure Captions**

**Figure 1**

(a) A schematic of a typical Si spin valve device (not to scale). The SOI substrate features a highly doped Si spin channel.

(b) Nonlocal spin signals from the Si spin valve device at room temperature under a dc bias current of +3.0 mA. An in-plane magnetic field is swept from -800 Oe to +800 Oe (red line) and from +800 Oe to -800 Oe (blue line).

(c) Hanle signals under a dc bias current of 3.0 mA, where the red and blue open circles denote Hanle signals under parallel and anti-parallel magnetization configurations. The red and blue lines are the fitting curves calculated by using the spin drift-diffusion equation.

**Figure 2**

(a) The measurement geometry for estimating the spin drift velocity. A perpendicular magnetic field was applied for Hanle measurements, and an electric field was applied between the NM and FM electrodes.

(b) Hanle signals at 8 K under dc bias conditions from -4.0 mA to -0.5 mA.

(c) The electric field dependence of the spin drift velocity at 8 K, as estimated from the analytical fitting function.

**Figure 3**

(a) A simulated curve of the relationship between the spin drift velocity and the ratio of the

spin resistance with and without spin drift.

(b) The measurement setup for the conventional NL-4T method.

(c) The measurement setup for the "crossed NL-4T" method. An electric field can be applied in the spin transport channel between two FM electrodes.

(d) The bias dependence of the spin signals at 8 K in the conventional and crossed NL-4T methods.

(e) Experimental results obtained by Hanle measurements with and without spin drift. Red and blue open circles denote the Hanle signals at 8 K under a bias current of, respectively, -4 mA and +4 mA in the crossed NL-4T method. The black open circles show the Hanle signals obtained by the conventional NL-4T method under a +4 mA bias current.

Simulated Hanle curves calculated by using the analytical solution, i.e., Eq. (3).

Figure 1

(a) 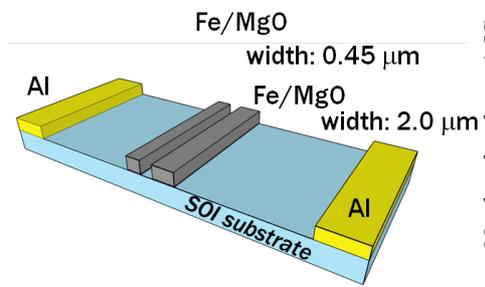 (b) 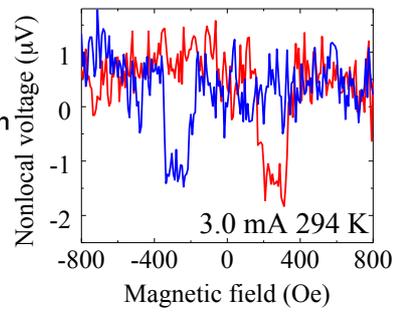 (c) 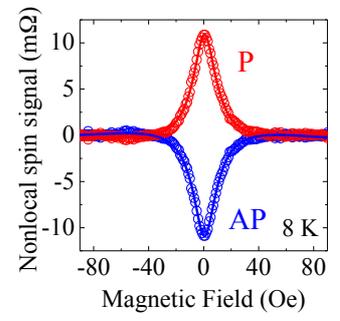

Figure 2

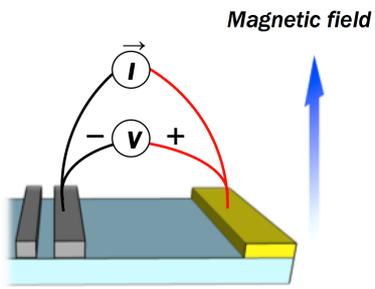
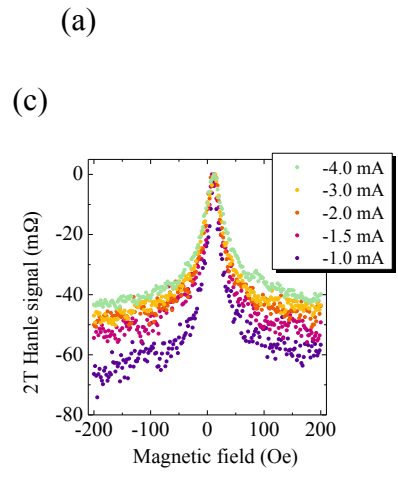
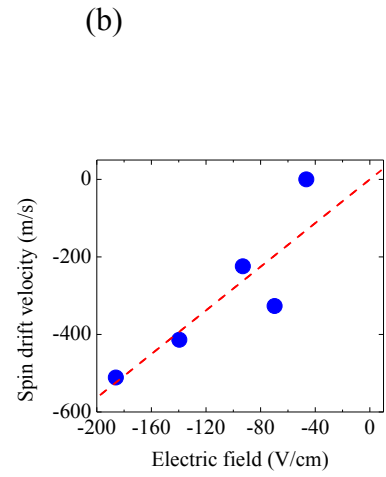

(a) (b) (c)

Figure 3

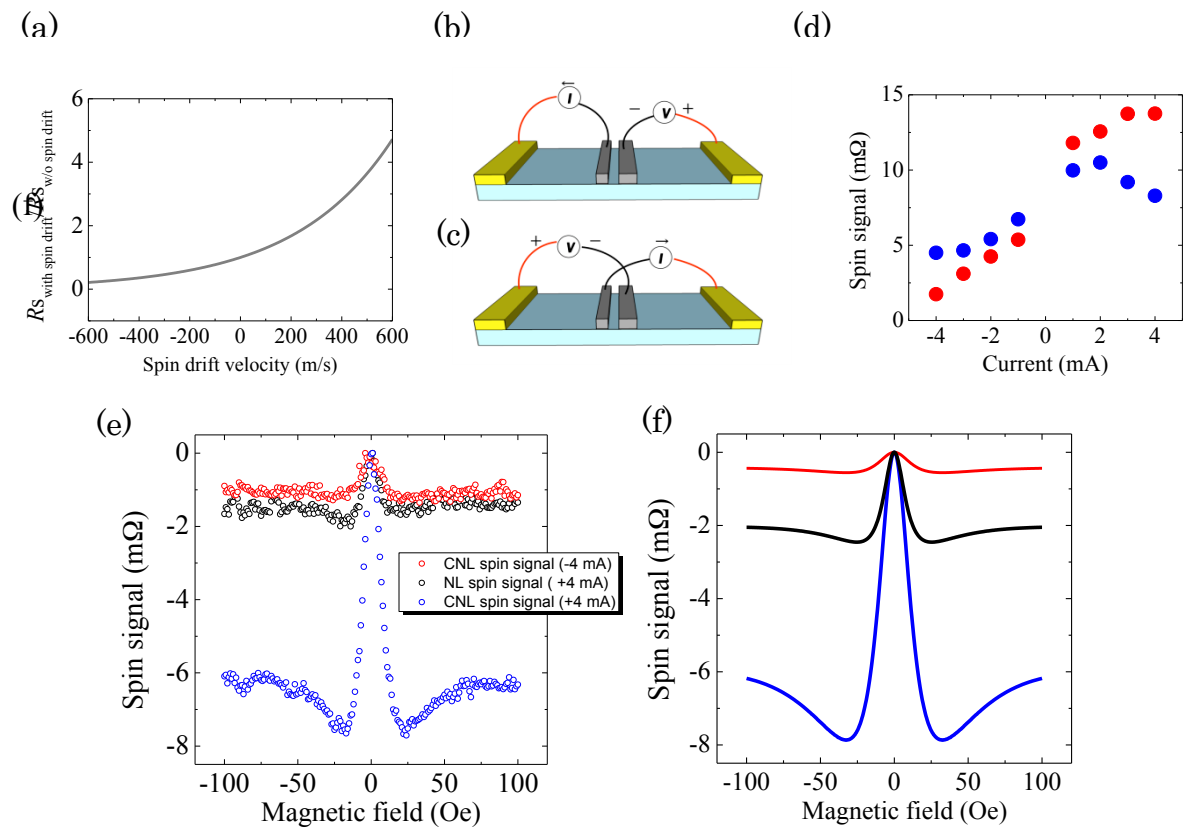